# Video Analysis and Modeling Tool for Physics Education: A workshop for Redesigning Pedagogy.

Loo Kang WEE[1], Tat Leong LEE[2]

[1]Ministry of Education, Education Technology Division, Singapore
[2]Ministry of Education, River Valley High School, Singapore
wee_loo_kang@moe.gov.sg, lee_tat_leong@moe.edu.sg

Abstract: This workshop aims to demonstrate how the Tracker Video Analysis and Modeling Tool engages, enables and empowers teachers to be learners so that we can be leaders in our teaching practice. Through this workshop, the kinematics of a falling ball and a projectile motion are explored using video analysis and in the later video modeling. We hope to lead and inspire other teachers by facilitating their experiences with this ICT–enabled video modeling pedagogy (Brown, 2008) and free tool for facilitating students-centered active learning, thus motivate students to be more self–directed.
Keyword: Tracker, active learning, education, teacher professional development, e–learning, open source, GCE Advance Level physics
PACS: 01.40.gb 01.50.H– 01.50.ht 01.50.hv 45.50.Dd

## I. INTRODUCTION

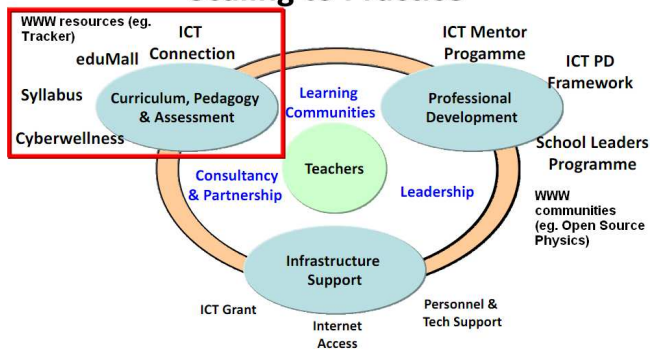

Figure 1. Scaling to Practice framework (MOE, 2010a, p. 20) adapted to include world wide web WWW resources like Tracker and global professional communities like Open Source Physics.

In support of the third masterplan for information and technology (ICT) in education vision of "Harnessing ICT, Transforming Learners" (MOE, 2009a, 2010a) and align with teacher leadership in professional development for pedagogical leaders (MOE, 2009b), this workshop aims to demonstrate how the Open Source Physics (Christian, 2010) community engages, enables and empowers (Romeo, 2006) teachers to be learners so that we can be leaders in our teaching practice. Workshops (Lee & Wee, 2011; Wee, Lee, & Chew, 2010; Wee & Tan, 2010a, 2010b), lesson examples in The ICT Connection portal (Goh et al., 2011; Lee, Wee, Cheng, & Tan, 2010) and mass briefing (Wee, 2010a) were conducted by the authors and others in a teacher-lead approach towards scaling of practice as in Figure 1 with meaningful use of Tracker as a pedagogical tool. The pedagogical strength of the Tracker program is it allow students to create simple dynamic particle model(s) on a video clip that we argue makes learning connected to real life and powerful as it provides a mechanism to progressively triangulate their understanding through the video model pedagogy (Brown, 2007, 2008, 2009, 2010).

Through this 75 minutes workshop, we aim to make aware to fellow physics educators the technological and pedagogical affordances provided by Tracker for supporting other teachers to scale this practice in their own classroom(s).

## II. INSTALATION, RUNNING OF TRACKER AND CONTENTS OF WORKSHOP

Tracker is a free video analysis and modeling tool built on the Open Source Physics (OSP) Java framework. The recommended installation method is to use the respective version 4.05 available at http://www.cabrillo.edu/~dbrown/Tracker/ depending on the operating system of the computer, though it is also possible to run from the Webstart. Readers may find this YouTube video (Wee, 2010b) that shows how to get the Tracker version 3.1, download and use it. There is a slight difference between the current version 4.05 and 3.1 but the YouTube video could still be useful.

All curriculum, professional development and ICT support materials of the workshop are available for download and use on the author's blog http://weelookang.blogspot.com/2011/05/video–analysis–and–modeling–tool–for.html (2011) for mass adoption and adaptation to scale this ICT enabled practice.

## III. KINEMATICS OF A FALLING BALL

In a free falling ball motion equation (1) represents its motion and equation (2) is the parabola fit equation for which the coefficients are equated, to derive the numerical values of the quantities such as acceleration in y direction, $a_y$ and initial velocity of motion in y direction, $u_y$.

$$y = \frac{1}{2}ay.t^2 + uy.t + 0 \qquad (1)$$

$$y = a*t^2 + b*t + c \qquad (2)$$

In Tracker's video analysis of the *y* vs *t* graph, after choosing the parabola fit of equation (2), the parameter are determine by the Tracker software as $a = -4.844$, $b = -0.28$ and $c = -0.001$. By comparing the equation (1) and (2), the students can infer the values of $\frac{1}{2}ay = -4.844$, thus $ay = -9.688$ $m/s^2$, $uy = -0.28$ $m/s$ and $c = -0.001$ $m$ respectively. The value $ay = -9.688$ $m/s^2$ can be interpreted to be approximately equal in value to the gravitational acceleration constant at the surface of Earth of $-9.81$ $m/s^2$ (3 significance figures, used in Advanced Level) and $-10$ $m/s^2$ (2 significance figures, used in Ordinary Level).

This activity is well suited to redesign ICT based pedagogy due to its ability to allow student to figure out the physics of motion through a real world video, without referring to the authoritative sources of knowledge such as teacher(s) and books.

A good teaching point that surfaced would be the appreciation of systematic error possible from the calibration stick error, and random error from the measurement or selection of the points of the path of the motion, experienced while in the act of being a scientist (Dewey, 1958).

## IV. PROJECTILE EQUATIONS AND ITS MATHEMATICAL DYNAMIC MODEL IN TRACKER

In an ideal projectile motion equation (3) and (4) represents the motion under constant acceleration.

$$\begin{pmatrix} vx \\ vy \end{pmatrix} = \begin{pmatrix} ux \\ uy \end{pmatrix} + \begin{pmatrix} ax \\ ay \end{pmatrix} t \qquad (3)$$

$$\begin{pmatrix} x \\ y \end{pmatrix} = \begin{pmatrix} ux \\ uy \end{pmatrix} t + \frac{1}{2} \begin{pmatrix} ax \\ ay \end{pmatrix} t^2 \qquad (4)$$

The dynamic particle model is selected as it is more suitable for this projectile motion instead of the analytic particle model as more complex drag force affected motion can be more easily modeled and be compared to the video. Readers may find this YouTube video (Wee, 2010c) useful that shows how the same process of building a dynamic model on the same projectile motion.

### A. Initial velocity in the x–direction & no x–direction force in projectile motion

Novice students generally may not fully appreciate the meaning of constant velocity in the x–direction of projectile motion as imply in equation (3x) and (4x) when $ax = 0$ $m/s^2$. We suggest an activity where student key in values for the initial velocity $vx$ in the dynamic model and observe the real data (red) versus the constant $vx$ model (pink), and make sense for themselves that instantaneous velocity is equal to 1.77 $m/s$ at all times of the projectile motion as in Figure 2.

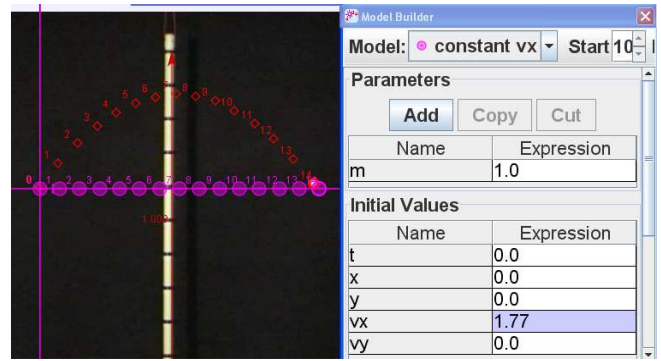

Figure 2. World view of a projectile motion with real data (red) versus the constant *vx* model (pink) on the left and the model builder of $vx = 1.77$ on the right.

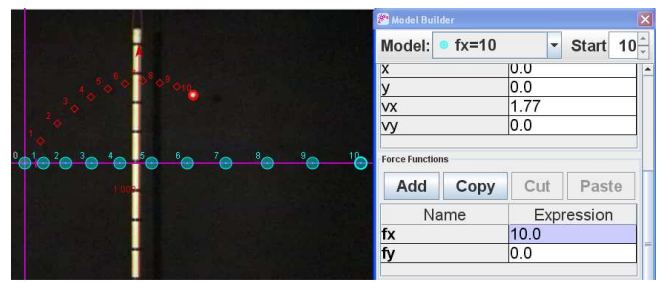

Figure 3. World view of a projectile motion with real data (red) versus the *fx* =10 N model (teal) on the left and the model builder values of $vx = 1.77$ and $fx = 10$ for this incorrect model on the right.

Novice students also have little means in typical classroom settings to understand why in projectile motion, there is no acceleration in the x direction where $ax = 0$ $m/s^2$, confused by their prior knowledge perhaps from movies of propelled projectile motion like rockets. Similarly, by keying in for the $fx \neq 0$ $N$ when mass of projectile $m = 1$ $kg$, the students can observe paths similar to Figure 3 for example $fx = 10$ $N$ and compare the real data (red) versus the $fx=10$ $N$ model (teal) to be not the vertically projected downward 'shadow' of the real data, thus this incorrect model is not representative of the real motion.

### B. The increment experience of video model building

We have used Tracker with our students and initial findings suggest that this kind of video modeling pedagogy is suitable for active and deep learning because the students can be said to be predicting by keying certain values, observing by compare the real data with the current proposed model, and explaining (White & Gunstone, 1992) by choice of values and linking to the video analysis data. Even with incorrect models input in, the results from the world view and associated multiple representational views (Wong, Sng, Ng, & Wee, 2011) in various scientific plots can allow the facilitation of data driven social discussions (Chai, Lim, So, & Cheah, 2011) among students and teacher(s). A possible completed model is shown in Figure 4 by trial and error, notice the 15[th] frame is slightly off from the real data but is somewhat close enough.

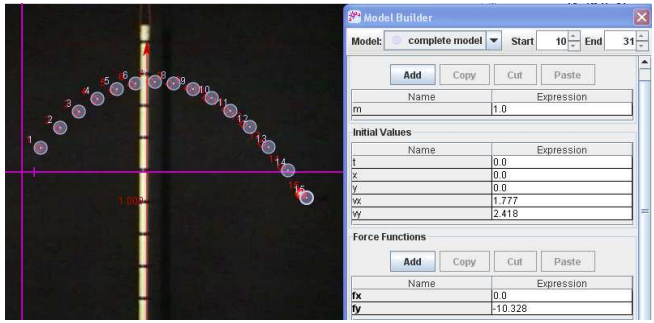

Figure 4. World view of a projectile motion with real data (red) versus the correct model of $vx = 1.777$ m/s, $vy = 2.418$ m/s and $fy = –10.328$ m/s$^2$ (light blue) by a largely trial and error approach to video modeling.

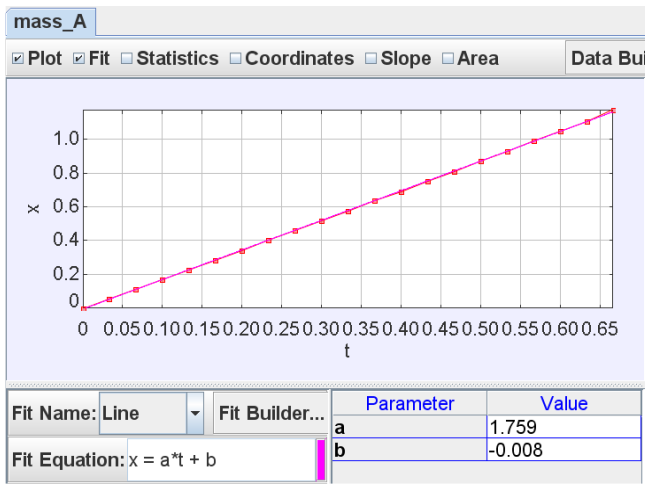

Figure 5. Data tool display real data mass_A of the $x$ versus $t$ view where a line fit equation of $x = a*t + b$ is used with parameter $a = 1.759$ and $b = –0.008$ as determined using Tracker. By comparing with equation (4x), it is determine that $vx = 1.759$ m/s.

## C. Using of analysis data to support choice of model building

Here is where we highlight to the students that the video analysis data of the same projectile motion (not detailed before) can be used to create a very accurate model. We found that students can have strong emotional responses as they didn't realize the analysis part was intended to allow for a means of data checking, that we hope can create a more lasting long-term memory effect on their learning. In our data analysis, we determine $vx = 1.759$ m/s as in Figure 5 and $ay = –10.124$ m/s$^2$ as well as $vy = 2.393$ m/s as in Figure 6.

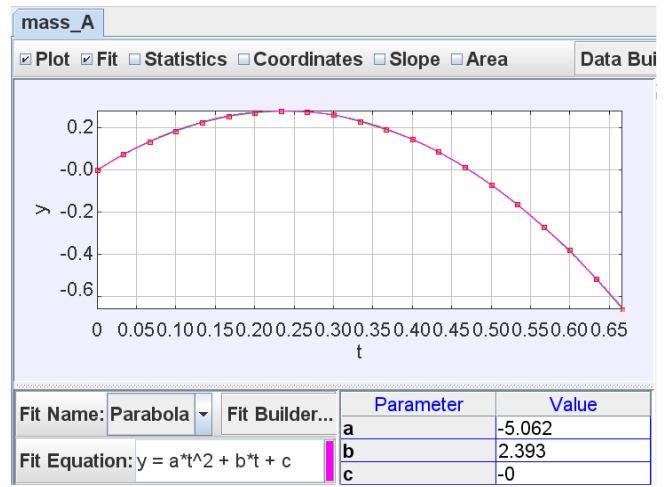

Figure 6. Data tool display real data mass_A $y$ versus $t$ view where parabola fit equation of $x = a^2*t + b*t + c$ is used with parameters $a = –5.062$, $b = –2.393$ and c = –0. By comparing with equation (4y), it is determine that $ay = –10.124$ m/s$^2$ and $vy = 2.393$ m/s respectively.

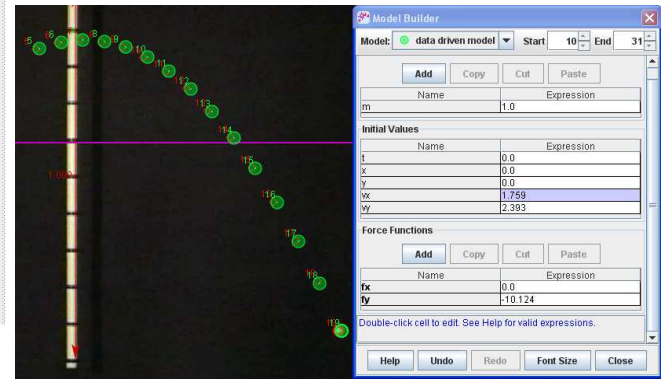

Figure 7. World view of a projectile motion with real data (red) versus the correct model of $vx = 1.759$ m/s, $vy = 2.393$ m/s and $fy = –10.124$ m/s$^2$ (green) by a video analysis data driven approach to video modeling.

As in our model, it is assumed that the mass $m =1$ kg, thus by Newton's Second Law in equation (5y), $fy = –10.124$ N.

$$\begin{pmatrix} fx \\ fy \end{pmatrix} = m \begin{pmatrix} ax \\ ay \end{pmatrix} \quad (5)$$

Figure 7 shows a more precise model with the values determined from video data analysis as the 19$^{th}$ frame shows the real data and data driven model to match very closely.

## D. Modeling $F = k*v$ and/or $k1*v^2$ to support the absence of air drag

Expert students can be challenged to extend their own learning (MOE, 2009c) to model air resistance and be convinced by the video model and real video that the video's projectile motion cannot be realistically assumed to be a motion with significant air resistance. The model for air resistance may be expressed as in equation (6) with both = $k*v$ and/or $k1*v^2$ depending on the model of air resistance equation assumed.

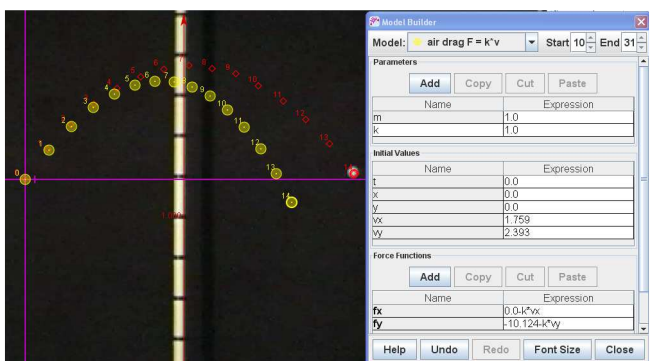

Figure 8. World view of a projectile motion with real data (red) versus the air drag model by *Fdrag* = *k*v* by inserting *fx* = 0–*k*vx* and *fy* = –10.124 – *k*vy* (yellow) on the left with the model builder values on the right.

$$\begin{pmatrix} fxdrag \\ fydrag \end{pmatrix} = k \begin{pmatrix} vx \\ vy \end{pmatrix} + k_1 \begin{pmatrix} vx^2 \\ vy^2 \end{pmatrix} \quad (6)$$

## V. CONCLUSION

The video data analysis of a falling ball motion is explored can allow students to self direct (Tan, Shanti, Tan, & Cheah, 2011) the learning and discover an accurate value for the gravitational constant of – 9.81 m/s$^2$.

This video modeling pedagogy is an active and fun way to understand physics of projectile motion that we view to have transform physics education (Weiman & Perkins, 2005). Four teaching and learning ideas in video modeling are highlighted and our initial research findings suggests this could be a deep pedagogical (MOE, 2010b) tool especially when video modeling (Brown, 2008) is combined with initial video analysis due to the data checking from the video model building process supported with the video analysis data.

## ACKNOWLEDGMENT

We wish to acknowledge Douglas Brown for the creation of this free tool Tracker and the Open Source Physics community mainly Francisco Esquembre, Fu–Kwun Hwang, Wolfgang Christian, Mario Belloni and Anne Cox and many others for their contributions in the development of Tracker and associated curriculum and resources.

## REFERENCE

Brown, D. (2007). *Combining computational physics with video analysis in Tracker*. Paper presented at the American Association of Physics Teachers AAPT Summer Meeting, Greensboro http://cabrillo.edu/~dbrown/tracker/air_resistance.pdf

Brown, D. (2008). *Video Modeling: Combining Dynamic Model Simulations with Traditional Video Analysis*. Paper presented at the American Association of Physics Teachers AAPT Summer Meeting, Edmonton. http://cabrillo.edu/~dbrown/tracker/video_modeling.pdf

Brown, D. (2009). *Video Modeling with Tracker*. Paper presented at the American Association of Physics Teachers AAPT Summer Meeting, Ann Arbor. http://cabrillo.edu/~dbrown/tracker/video_modeling.pdf

Brown, D. (2010). Tracker Introduction to Video Modeling (AAPT 2010). Portland Oregon. Retrieved from http://www.compadre.org/Repository/document/ServeFile.cfm?ID=10188&DocID=1749

Chai, C. S., Lim, W.-Y., So, H.-J., & Cheah, H. M. (2011). Advancing collaborative learning with ICT: Conception, Cases and Design. MOE (Ed.) Retrieved from http://ictconnection.edumall.sg/ictconnection/slot/u200/mp3/monographs/advancing%20collaborative%20learning%20with%20ict.pdf

Christian, W. (2010). Open Source Physics (OSP) Retrieved 25 August, 2010, from http://www.compadre.org/osp/

Dewey, J. (1958). *Experience and nature*: Dover Pubns.

Goh, J., Wee, L. K., Leong, T. K., Bakar, R. A., Koh, J. M., Tan, H. K., & Tan, E. (2011). Learning Physics of Projectile through Video Analysis and Modeling Retrieved 02 June, 2010, from http://ictconnection.edumall.sg/cos/o.x?ptid=711&c=/ictconnection/ictlib&func=view&rid=533

Lee, T. L., & Wee, L. K. (2011). Workshop on investigation of the kinematics of a falling ball through Video Analysis and Modeling. *3rd Instructional Program Support Group (IPSG) Physics* Retrieved 19 January, 2011, from http://weelookang.blogspot.com/2011/05/workshop-on-learning-physics-of-sport.html

Lee, T. L., Wee, L. K., Cheng, S. S. S., & Tan, Y. L. (2010). Learning Physics of Sport Science through Video Analysis and Modeling Retrieved 02 June, 2010, from http://ictconnection.edumall.sg/cos/o.x?ptid=711&c=/ictconnection/ictlib&func=view&rid=82

MOE. (2009a). Speech by Mr S Iswaran, Senior Minister of State, Ministry of Trade and Industry and Ministry of Education, at the International Conference on Teaching and Learning with Technology (iCTLT) on Thursday, 4 March 2010, at 9.00am at Suntec Singapore International Convention and Exhibition Centre Retrieved 20 October, 2010, from http://www.moe.gov.sg/media/speeches/2010/03/04/speech-by-mr-s-iswaran-at-ictlt-2010.php

MOE. (2009b). Teachers — The Heart of Quality Education Retrieved 20 October, 2010, from http://www.moe.gov.sg/media/press/2009/09/teachers-the-heart-of-quality.php

MOE. (2009c). Third Masterplan for ICT in Education Retrieved 20 October, 2010, from http://ictconnection.edumall.sg/cos/o.x?c=/ictconnection/pagetree&func=view&rid=665

MOE. (2010a). Harnessing ICT Transforming Learners NIE TE21 Summit 02 Nov 2010 Masterplan for ICT in Education Retrieved 20 June, 2011, from http://www.nie.edu.sg/files/TE21%20Summit_ETD%20slides_finalv2.pdf

MOE. (2010b). An Introduction to PLCs Retrieved 01 December, 2010, from http://www.academyofsingaporeteachers.moe.gov.sg/cos/o.x?c=/ast/pagetree&func=view&rid=1069395

Romeo, G. (2006). Engage, Empower, Enable: Developing a Shared Vision for Technology in Education. In D. Hung & M. S. Khine (Eds.), *Engaged Learning with Emerging Technologies* (pp. 149-175): Springer Netherlands.

Tan, S. C., Shanti, D., Tan, L., & Cheah, H. M. (2011). Self-directed learning with ICT: Theory, Practice and Assessment. MOE (Ed.) Retrieved from http://ictconnection.edumall.sg/ictconnection/slot/u200/mp3/monographs/self-directed%20learning%20with%20ict.pdf

Wee, L. K. (2010a, 03 November). eduLab mass briefing on possible ideation options for eduLab projects sharing on Easy Java Simulation and Tracker. *Jurong Junior College*, 2010, from http://weelookang.blogspot.com/2010/10/edulab-mass-briefing-at-jurong-junior.html

Wee, L. K. (Producer). (2010b). Tracker Free Video Analysis for Physics Education by Douglas Brown & Youtube PD by lookang. Retrieved from http://www.youtube.com/watch?v=cuYJsnhWXOw

Wee, L. K. (Producer). (2010c). Tracker Free Video Modeling for Physics Education by Douglas Brown & Youtube PD by lookang. Retrieved from http://www.youtube.com/watch?v=WSG1x3klkH0

Wee, L. K., & Lee, T. L. (2011). *Video Analysis and Modeling Tool for Physics Education*. Paper presented at the 4th Redesigning Pedagogy International Conference, Singapore. http://weelookang.blogspot.com/2011/05/video-analysis-and-modeling-tool-for.html

Wee, L. K., Lee, T. L., & Chew, C. (2010, 23-24 November). *Workshop Concurrent 4.9 Workshop - Innovation in Science Education Open Source Physics – Tracker Video Analysis and Modeling Tool*. Paper presented at the Singapore 1st Science Teacher Conference, Singapore.

Wee, L. K., & Tan, S. (2010a, 20 May). ICT mentor sharing secondary science on video analysis using Tracker. *River Valley High School*, 2010, from http://ictconnection.edumall.sg/cos/o.x?ptid=709&c=/ictconnection/forum&func=showthread&t=64

Wee, L. K., & Tan, S. (2010b, 14 May). ICT mentor sharing secondary science on video analysis using Tracker. *Queensway Secondary School*, 2010, from http://ictconnection.edumall.sg/cos/o.x?ptid=709&c=/ictconnection/forum&func=showthread&t=64

Weiman, C., & Perkins, K. (2005). Transforming Physics Education. *Physics Today, 58*(11), 36-40.

White, R. T., & Gunstone, R. F. (1992). *Probing understanding*: Routledge.

Wong, D., Sng, P. P., Ng, E. H., & Wee, L. K. (2011). Learning with multiple representations: an example of a revision lesson in mechanics. *Physics Education, 46*(2), 178.

## AUTHOR

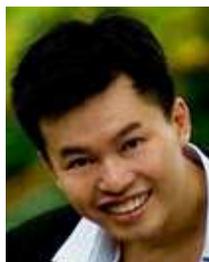

Loo Kang WEE is currently an educational technology officer at the Ministry of Education, Singapore and a PhD candidate at the National Institute of Education, Singapore. He was a junior college physics lecturer and his research interest is in designing simulations for physics learning. His curriculum material and simulation models can be accessed from the NTNUJAVA Virtual Physics Laboratory and Open Source Physics Collection under Creative Commons Attribution License.

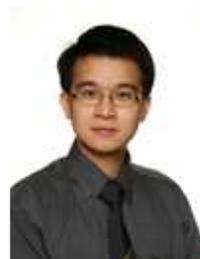

Tat Leong LEE is Head of Department for Information and Technology at River Valley High School and a Physics teacher.